\documentclass[12pt]{spieman}  
\usepackage{amsmath,amsfonts,amssymb}
\usepackage{graphicx}
\usepackage{setspace}
\usepackage{tocloft}

\title{Power of simultaneous X-ray and UV high-resolution spectroscopy for probing AGN outflows}

\author[a,*]{Missagh Mehdipour}
\author[b]{Laura W. Brenneman}
\author[c]{Jon M. Miller}
\author[d,e]{Elisa Costantini}
\author[f,g]{\\Ehud Behar}
\author[h]{Luigi C. Gallo}
\author[d,i]{Jelle S. Kaastra}
\author[j,k,l]{Sibasish Laha}
\author[m]{Michael A. Nowak}

\affil[a]{Space Telescope Science Institute, 3700 San Martin Drive, Baltimore, MD 21218, USA}
\affil[b]{Center for Astrophysics, Harvard \& Smithsonian, 60 Garden Street, Cambridge, MA 02138, USA}
\affil[c]{Department of Astronomy, University of Michigan, 1085 South University Avenue, Ann Arbor, MI 48109, USA}
\affil[d]{SRON Netherlands Institute for Space Research, Niels Bohrweg 4, 2333 CA Leiden, the Netherlands}
\affil[e]{Anton Pannekoek Institute, University of Amsterdam, Postbus 94249, 1090 GE Amsterdam, The Netherlands}
\affil[f]{Department of Physics, Technion, Haifa 32000, Israel}
\affil[g]{MIT Kavli Institute for Astrophysics and Space Research, Massachusetts Institute of Technology, Cambridge, MA 02139, USA}
\affil[h]{Department of Astronomy \& Physics, Saint Mary’s University, 923 Robie Street, Halifax, NS B3H 3C3, Canada}
\affil[i]{Leiden Observatory, Leiden University, PO Box 9513, 2300 RA Leiden, the Netherlands}
\affil[j]{Center for Space Science and Technology, University of Maryland Baltimore County, 1000 Hilltop Circle, Baltimore, MD 21250, USA}
\affil[k]{Astrophysics Science Division, NASA Goddard Space Flight Center, Greenbelt, MD 20771, USA}
\affil[l]{Center for Research and Exploration in Space Science and Technology, NASA/GSFC, Greenbelt, Maryland 20771, USA}
\affil[m]{Department of Physics, Washington University in St. Louis, Campus Box 1105, One Brookings Drive, St. Louis, MO 63130-4899, USA}

\cftpagenumbersoff{figure}
\cftpagenumbersoff{table}
\usepackage{xspace}
\usepackage{aas_macros}
\newcommand{\NH}{\ensuremath{N_{\mathrm{H}}}\xspace}

\usepackage[T1]{fontenc}

\DeclareRobustCommand{\ion}[2]{\textup{#1\,\textsc{\lowercase{#2}}}}
\newcommand*\element[1][]{%
  \def\aa@element@tr{#1}%
  \aa@element}

\begin{document} 
\maketitle

\begin{abstract}
Black hole accretion in active galactic nuclei (AGN) is coupled to the evolution of their host galaxies. 
Outflowing winds in AGN can play an important role in this evolution through the resulting feedback mechanism. 
Multi-wavelength spectroscopy is key for probing the intertwined physics of inflows and outflows in AGN.
However, with the current spectrometers, crucial properties of the ionized outflows are poorly understood, such as their coupling to the accretion rate, their launching mechanism, and their kinetic power.
In this paper we discuss the need for simultaneous X-ray and UV high-resolution spectroscopy for tackling outstanding questions on these outflows in AGN.
The instrumental requirements for achieving the scientific objectives are addressed. We demonstrate that these requirements would be facilitated by the proposed Arcus Probe mission concept.
The multi-wavelength spectroscopy and timing by Arcus would enable us to establish the kinematics and ionization structure of the entire ionized outflow, extending from the vicinity of the accretion disk to the outskirts of the host galaxy. 
Arcus would provide key diagnostics on the origin, driving mechanism, and the energetics of the outflows, which are useful benchmarks for testing various theoretical models of outflows and understanding their impact in AGN. 

\end{abstract}

\keywords{spectroscopy, active galactic nuclei, outflows, accretion disk}

{\noindent \footnotesize\textbf{*}Missagh Mehdipour, \linkable{mmehdipour@stsci.edu}}

\begin{spacing}{1}

\section{Introduction}
\label{sect_intro}

Outflows/winds in AGN transport mass and energy away from the central engine and into the host galaxy. Ascertaining the physical structure and energetics of these outflows, and understanding how they are launched and driven, are important for determining their role in AGN feedback.  
Currently, the dynamics, kinematics, and ionization structure of the ionized outflows are poorly understood. This makes it challenging to determine how their momentum and energy propagate into the galaxy, and how they impact their environment. Different types/forms of ionized outflows, with distinct characteristics, have been observed at the micro (sub-pc) scale (disk and the broad-line region, BLR), the meso (pc) scale (the torus and the narrow-line region, NLR), and the macro (kpc) scale (the host galaxy environment) \cite{Laha21,Gall23}. The formation of these various ionized outflows, and their relation to each other, are currently not well understood. The origin (disk or torus) and driving mechanism (thermal, radiative, or magnetic) of the outflows remain open questions. Deriving the parameters of AGN winds from multi-wavelength observations is needed in order to investigate their origin and launching mechanism\cite{Fuku24}. 

The proposed {\it Arcus Probe} mission \cite{Smith24} would help us in solving these outstanding questions on AGN outflows. The simultaneous high-resolution spectroscopy with the X-ray Spectrometer (XRS) and UV Spectrometer (UVS) of Arcus would be instrumental in probing the currently poorly-understood properties of AGN outflows. In the following section we discuss how simultaneous X-ray and UV high-resolution spectra enable us to advance our understanding of AGN outflows. We present the instrumental requirements in Sect. \ref{sect_merit} and compare the capabilities of Arcus with existing missions. Next, in Sect. \ref{sect_timing}, we show how spectral timing of absorption lines measured by Arcus helps us to derive the sought-after density, location, and energetics of the various components of ionized outflows. Finally, in Sect. \ref{sect_redshift} we discuss how Arcus enables us to extend high-resolution X-ray/UV spectroscopy of outflows beyond the local universe to higher redshift AGN.

\section{Necessity for simultaneous X-ray and UV high-resolution spectroscopy: Arcus Probe}
\label{sect_simult}

Outflows of gas in AGN are photoionized by the accretion-powered continuum radiation, referred to as the spectral energy distribution (SED). 
These ionized outflows imprint their spectral signatures on the SED, predominantly as absorption lines in the soft X-ray and far-UV energy bands.
High-resolution spectroscopy is required to resolve and model individual spectral lines from the outflows, and thereby to investigate their physical parameters. 
The ionization state of the outflowing gas is quantified by its ionization parameter $\xi$ \cite{Tar69}, defined as
\begin{equation}
\label{eq_xi}
\xi \equiv \frac{L_{\rm ion}}{{n_{\rm{H}}\,r^2 }}
\end{equation}
where $L_{\rm ion}$ is the luminosity of the ionizing source over 1--1000 Ryd (13.6 eV to 13.6 keV) in $\rm{erg}\ \rm{s}^{-1}$, $n_{\rm{H}}$ the hydrogen density in $\rm{cm}^{-3}$, and $r$ the distance between the gas and ionizing source in $\rm{cm}$.
Depending on the parameters of the outflow, particularly its $\xi$, the absorption lines may appear in either X-ray, UV, or both bands.

To obtain the ionization structure of the outflows, as well as their kinematic structure, all types and regions of the outflows from near the disk/BLR, to the NLR and the host galaxy environment, are needed to be captured.
In order to achieve this, joint X-ray and UV high-resolution spectroscopy is required. Importantly, since the parameters of the outflows can vary over time, the joint X-ray and UV data are needed to be taken simultaneously.
This variability may be induced by changes in the flux or shape of the ionizing SED, thereby causing changes in the ionization state of the gas, hence resulting in changes in the strength of the X-ray and UV absorption lines.
Additionally, there can be intrinsic changes in the parameters of the wind, such as its column density \NH changing as it crosses our line of sight, or the wind evolving over time, again resulting in variability in the X-ray and UV absorption lines.
Thus, simultaneity is an important requirement to study these variable phenomena.
Furthermore, the simultaneous X-ray and UV exposure helps in the construction of the broadband ionizing SED, which is needed for photoionization modeling of the outflows.
Therefore, to model the components of ionized outflows, and correctly decipher the nature of their spectral line variability, the X-ray and UV spectra need to be taken simultaneously. 
This is a crucial capability of the Arcus mission.

\begin{figure*}
\resizebox{1.0\hsize}{!}{\includegraphics[angle=0]{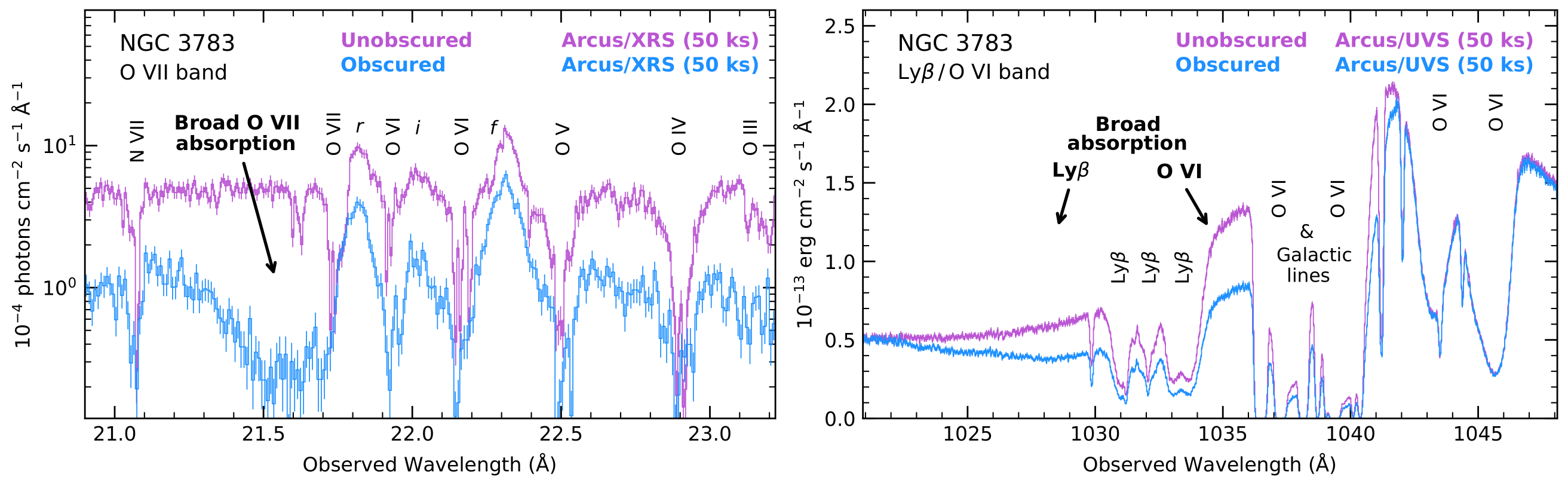}}
\caption{\small Simulated Arcus XRS (left panel) and UVS (right panel) spectra of NGC 3783 taken in its unobscured and obscured states. There are narrow and broad absorption lines in both the X-ray and UV bands from AGN winds. Useful variabilities in these absorption lines occur on timescales of hours and days. Simultaneous X-ray and UV spectroscopy with Arcus allows us to trace and utilize these variabilities to obtain the poorly-understood properties of the AGN winds: density, location, and hence their energetics.
\label{fig_spec}}
\vspace{-0.0cm}
\end{figure*}

We use the case of the Seyfert-1 galaxy NGC 3783 to demonstrate the necessity for simultaneous X-ray and UV high-resolution spectroscopy. 
The simulated spectra with the XRS and UVS of Arcus are shown in Fig. \ref{fig_spec}.
The models used for the simulations are based on our previous studies of NGC 3783 with XMM-Newton \cite{Mehd17,Cost22}, HST/COS \cite{Kris19a}, and FUSE \cite{Gab03}.
The {\tt SPEX} v3.07.01 package \cite{Kaa96,Kaas20} and its {\tt pion} \cite{Meh16b,Mill15} photoionization model are used in our simulations.
The figure shows the presence of narrow absorption lines from a ``warm-absorber" outflow, and broad absorption lines from an obscuring disk wind. 
These outflows are associated with the NLR and BLR, respectively.
Their spectral lines are shown in both the X-ray (\ion{O}{VII}) and far-UV (\ion{O}{VI} / Ly$\beta$) bands.
There is crucial variability in these X-ray and UV lines, which can only be understood by simultaneous high-resolution spectroscopy and modeling.
The transient obscuring disk wind appears in our line of sight in the BLR, producing broad X-ray and UV absorption lines \cite{Mehd17}.
This inner obscuration causes gas located further out in the NLR to become less ionized, resulting in changes in their narrow absorption lines.

%

%
\begin{figure*}
\centering
\resizebox{0.65\hsize}{!}{\includegraphics[angle=0]{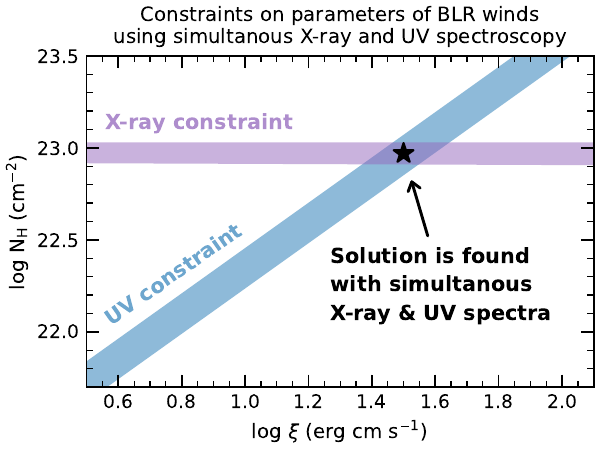}}
\caption{\small An example from NGC 5548, demonstrating why joint X-ray and UV spectroscopy and photoionization modeling of AGN winds is important. In order to obtain a unique photoionization solution for the BLR/disk winds, assimilation of spectral information from both X-rays and UV is needed. As BLR/disk winds are highly variable and transient in nature, the X-ray and UV spectra need to be taken simultaneously. This becomes achievable with the Arcus mission.
\label{fig_BLR}}
\vspace{-0.0cm}
\end{figure*}

Lack of simultaneous X-ray and UV spectra leaves ambiguity in the modeling and interpretation of AGN winds, which is usually the case when pairing current X-ray missions with HST.
It is challenging to coordinate simultaneous observations, and even in these rare cases, the much shorter HST exposure overlaps with a tiny portion of the much longer X-ray exposure.
It is not feasible to obtain few tens/hundreds of ks with HST to overlap with a single X-ray observation. 
Furthermore, target-of-opportunity (ToO) observations with HTS have constraints that make it challenging to coordinate simultaneous ToO observations with the X-ray missions, especially at response times of few days or less.
The Arcus Probe mission is designed to overcome these limitations and difficulties, by providing simultaneous X-ray and UV spectral coverage, with high signal-to-noise (S/N) and high energy resolution, within 24 hours.

Finally, Fig. \ref{fig_BLR} illustrates why simultaneous X-ray and UV spectral modeling is essential for studying AGN winds. This is shown for the case of a strong obscuring wind, such as the one seen in the Seyfert-1 galaxy NGC 5548 \cite{Kaas14,Kris19b,Mehd22c}. The high column density, and low X-ray covering fraction of such BLR winds, causes the ionization parameter $\xi$ to be unconstrained with the X-rays alone. On the other hand, because the UV continuum is mostly unaffected by the obscuration, the modeling of the UV absorption lines allows the ionization parameter to be estimated. However, because of UV line saturation, one cannot constrain the total column density \NH of the gas, whereas it can be measured from modeling the X-ray continuum absorption. Therefore, only by combining these X-ray and UV spectroscopic information a unique photoionization solution for the wind is found.
Furthermore, multi-wavelength spectroscopy is needed for investigating the ionization and thermal structure of AGN winds, and hence examining the thermal instability within the wind. This thermal instability can result in clumpiness and partially covering of the wind over time\cite{Mehd22c,Prog22,Wate22}, which needs to be understood for properly interpreting the measured parameters and linking observations to the theoretical models of the winds.

\section{Instrumental requirements for spectral line diagnostics of AGN outflows: comparison of Arcus with existing missions}
\label{sect_merit}

\begin{figure*}
\centering
\resizebox{1.0\hsize}{!}{\includegraphics[angle=0]{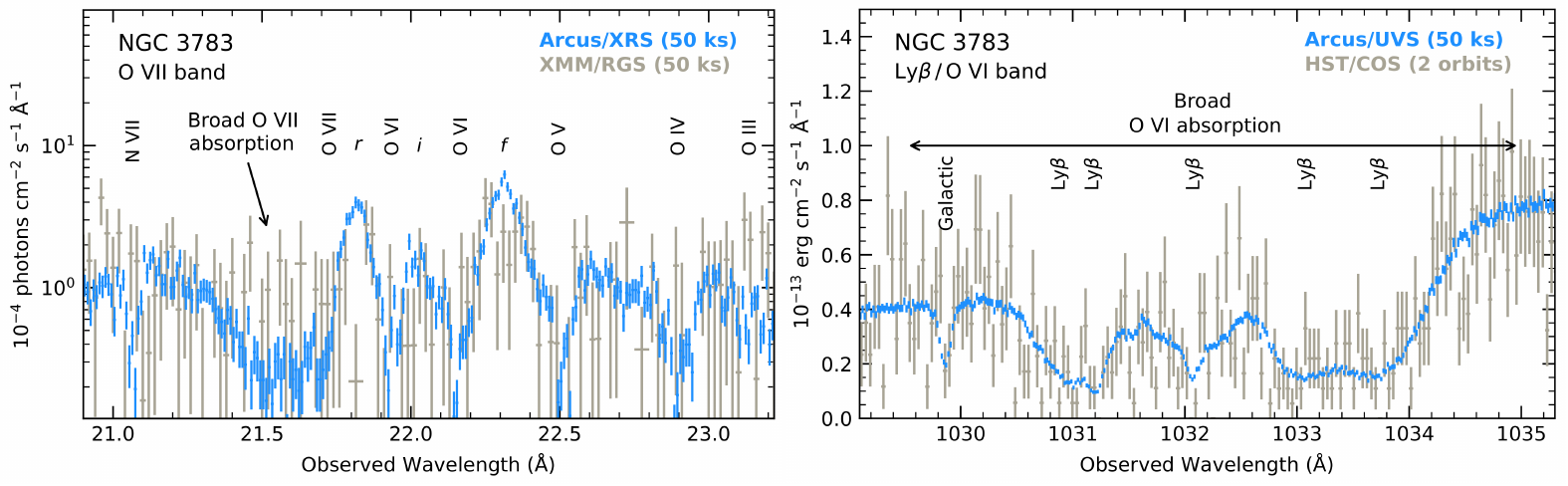}}
\caption{\small Comparison of simulated Arcus/XRS with XMM/RGS spectra (left panel), and Arcus/UVS with HST/COS spectra (right panel) for the obscured state of NGC 3783. With existing missions it is not feasible to measure the individual absorption lines and trace their variability, resulting in large uncertainties in the parameterization of AGN winds (Fig. \ref{fig_density} and \ref{fig_redshift}), thus making our interpretations ambiguous. This limitation is overcome by the Arcus mission.
\label{fig_compare}}
\vspace{-0.0cm}
\end{figure*}

Figure \ref{fig_compare} shows simulated spectra of NGC 3783, comparing Arcus/XRS with XMM/RGS in the O VII band, and Arcus/UVS with HST/COS in the O VI / Ly$\beta$ band.
The selected exposure times are based on typical ToO observations of transient obscuring winds: 50~ks with XMM and two orbits with HST/COS.
As seen in Fig. \ref{fig_compare}, Arcus makes it possible to easily resolve and detect individual absorption lines, whereas this is too difficult with existing missions.
To quantify and compare the capabilities of different instruments for line diagnostics, we make use of the following figures of merit (FoMs).
For more details and discussion on these FoMs we refer to \cite{Kaas08b}.

%

%
For line detection, the FoM $M_{l}$ is defined as
\begin{equation}
\label{eq:M_l}
M_{l} \equiv \dfrac{S}{\sqrt{F_{l}\, t}}
\end{equation}
where, $S$ is the S/N ratio, $F_{l}$ the flux of the line, and $t$ the exposure time of the observation. 
Using Poisson statistics, ${S = N_{l} / \sqrt{N}}$, where $N_{l}$ is the number of counts in the line, and ${N = N_{c} + N_{b} + N_{l}}$ is the total number of counts in the bin, including that of the continuum ($N_{c}$) and background ($N_{b}$). The line counts ${N_{l} = F_{l}\,t\,A}$, where $A$ is the effective area of the instrument. Similarly, ${N_{c} = F_{c}\,t\,A\,\Delta E}$, and ${N_{b} = F_{b}\,t\,A\,\Delta E}$, where ${\Delta E}$ is the resolution element. Therefore, S/N ratio $S$ can be expressed as
\begin{equation}
\label{eq:S}
S = \dfrac{\sqrt{F_{l}\,t\,A}}{\sqrt{1 + \dfrac{\Delta E}{W}\left(1 + \dfrac{F_{b}}{F_{c}}\right)}}
\end{equation}
where the line equivalent width ${W \equiv F_{l} / F_{c}}$. By substituting into Eq. \ref{eq:M_l} we get
\begin{equation}
M_{l} \propto
\begin{cases}
\sqrt{A\,}        & \mathrm{(strong\ line)}\\
\sqrt{A/\Delta E} & \mathrm{(weak\ line)} 
\end{cases}
\end{equation}

\noindent Next, we define the FoM for measuring the line centroid, $M_{v}$, as
\begin{equation}
\label{eq:M_v}
M_{v} \equiv \dfrac{c}{\Delta v}
\end{equation}
where, $c$ is the speed of light, and $\Delta v$ the measurement accuracy of the line centroid in km~s$^{-1}$. For an instrument with a Gaussian redistribution function with r.m.s. width ${\sigma_{0} = \Delta E / \sqrt{\ln 256}}$, and thus ${\Delta v = c\, \Delta E / \sqrt{\ln 256}\, E\, S}$, by substituting into Eq. \ref{eq:M_v} we obtain
\begin{equation}
M_{v} \propto
\begin{cases}
(E/\Delta E)\, \sqrt{A}        & \mathrm{(strong\ line)}\\
(E/(\Delta E)^{1.5})\, \sqrt{A}        & \mathrm{(weak\ line)} 
\end{cases}
\end{equation}

\noindent The FoM for velocity broadening is defined as 
\begin{equation}
\label{eq:M_sigma}
M_{\sigma} \equiv \dfrac{1}{\Delta \sigma_{v}}
\end{equation}
where, $\Delta \sigma_{v}$ is the uncertainty in the r.m.s. width in km~s$^{-1}$. For an instrument with Gaussian redistribution, $\Delta \sigma_{v}$ can be expressed as
\begin{equation}
\label{eq:sigma_v}
\Delta \sigma_{v} = \dfrac{c^{2}\,(\Delta E)^{2}}{\sqrt{2}\,\ln 256\,\sigma_{v}\,E^{2}\,S}
\end{equation}
Substituting Eqs. \ref{eq:S} and \ref{eq:sigma_v} into Eq. \ref{eq:M_sigma} gives
\begin{equation}
M_{\sigma} \propto
\begin{cases}
(E^{2}/(\Delta E)^2)\, \sqrt{A}            & \mathrm{(strong\ line)}\\
(E^{2}/(\Delta E)^{2.5})\, \sqrt{A}        & \mathrm{(weak\ line)} 
\end{cases}
\end{equation}

\noindent Finally, the figure of merit for an unresolved line structure is defined as
\begin{equation}
\label{eq:M_u}
M_{u} \equiv \dfrac{E}{\Delta E}
\end{equation}

In Table \ref{table_merits} we list the FoMs of Arcus/XRS with XMM/RGS, and those of Arcus/UVS with HST/COS. These comparisons are carried at X-ray wavelength of 22 \AA\ (\ion{O}{VII} line) and the UV wavelength of 1032 \AA\ (\ion{O}{VI} line).
The superior effective area $A$ and energy resolution ${\Delta E}$ of Arcus ensures that lines from various outflow components, as well as any blended foreground Milky Way lines, can be properly disentangled and parameterized.
The improvement offered by Arcus relative to the other missions is given by the ratio of FoMs in Table \ref{table_merits}. 
In the following Sects. \ref{sect_timing} and \ref{sect_redshift}, the impact of these instrumental improvements by Arcus are seen in the results of our simulations for measuring key parameters of AGN outflows.

%
\begin{table}[!t]
\caption{Figures of merit (FoMs) of Arcus compared with those of XMM/RGS and HST/COS. They correspond to detection of strong and narrow absorption lines at 22 \AA\ (region of \ion{O}{VII} line) and 1032 \AA\ (region of \ion{O}{VI} line). The effective area $A$ is given in cm$^{-2}$ and the energy resolution $\Delta E$ in eV.} 
\vspace{-0.2cm}
\label{table_merits}
\begin{center}
\setlength{\tabcolsep}{5.4pt}
\small
\begin{tabular}{c | c c | c c | c c | c c | c}
\hline
\rule[-1ex]{0pt}{3.5ex} Instrument & $A$            & $\Delta E$ &  \multicolumn{2}{c|}{$M_{l}$}  &   \multicolumn{2}{c|}{$M_{v}$}  &  \multicolumn{2}{c|}{$M_{\sigma}$}  &  $M_{u}$ \\
\rule[-1ex]{0pt}{3.5ex}            &                &            &  Strong &   Weak              &  Strong  & Weak                &   Strong    & Weak                 &          \\
\hline\hline
\rule[-1ex]{0pt}{3.5ex} Arcus XRS   & 450 & 0.16   & 21.2 & 53.0 & 76,100 & 190,260 & ${2.73 \times 10^{8}}$  & ${6.83 \times 10^{8}}$ & 3588 \\
\rule[-1ex]{0pt}{3.5ex} XMM RGS     & 42    & 1.79   & 6.5  & 4.8  & 2080  & 1550   & 666,400 & 498,100 & 321 \\
\rule[-1ex]{0pt}{3.5ex} XRS $\div$ RGS & 10.7 & 0.089 & 3.3  & 11.0 & 36.6   & 123     & 410     & 1370    & 11.2    \\
\hline
\rule[-1ex]{0pt}{3.5ex} Arcus UVS   & 445   & 0.042  & 21.1 & 103  &  6040 & 29,450  & ${1.73 \times 10^{6}}$  & ${8.42 \times 10^{6}}$ & 286 \\
\rule[-1ex]{0pt}{3.5ex} HST COS     & 25    & 0.137  & 5.0  & 13.5 &  440   & 1190   & 38,460   & 103,900  & 87.7 \\
\rule[-1ex]{0pt}{3.5ex} UVS $\div$ COS & 17.8 & 0.31  & 4.2  & 7.6  &  13.7  & 24.7    & 45.0     & 81.0     & 3.3  \\
\hline 
\end{tabular}
\end{center}
\end{table} 

\section{Spectral timing technique to further probe uncertain parameters of AGN outflows: breakthrough with Arcus}
\label{sect_timing}

Spectral variability is a useful characteristic of AGN for probing the unknown properties of outflows, namely their density $n_{\rm H}$ and location.
Importantly, these are needed to calculate the kinetic power of the outflows, $P_{kin}$, which is subsequently used to assess the impact of the outflows on their environment.
The kinetic power 
\begin{equation}
\label{eq_power}
P_{\rm kin} \propto N_{\rm H}\,v_{\rm out}^3\,r
\end{equation}
where, $N_{\rm H}$ is the total column density, $v_{\rm out}$ the outflow velocity, and $r$ the distance of the gas from the central ionizing source.
The $N_{\rm H}$ and $v_{\rm out}$ are measured directly from spectroscopy, while $r$ remains unknown.  
From photoionization modeling of the ions that are present in the spectra, the ionization parameter $\xi$ (Eq. \ref{eq_xi}) is determined.
And from broadband continuum modeling, using simultaneous UV and X-ray data, $L_{\rm ion}$ is obtained.
Therefore, if we obtain $n_{\rm H}$, Eq. \ref{eq_xi} would provide $r$, and hence $P_{\rm kin}$ of the gas can be calculated using Eq. \ref{eq_power}

The recombination timing technique uses the response of the gas to the variability of the ionizing SED to measure the density $n_{\rm H}$.
The recombination time $t_{\rm rec}$ of an ion recombining from ionization state ${i+1}$ to $i$ is ${t_{\rm rec} =  n_{i}\, /\, (n_{i+1}\, \alpha_{i+1}\, n_{\rm e} - n_{i}\, \alpha_{i}\, n_{\rm e})}$, where $n_{\rm e}$ is the electron density, ${n_{i+1}}$ and ${n_{i}}$ are the densities of the ions in state ${i+1}$ and $i$, respectively, and ${\alpha_{i+1}}$ and ${\alpha_{i}}$ are the recombination rate coefficients (${{\rm cm}^{3}~{\rm s}^{-1}}$) from state ${i + 1}$ to $i$, and ${i}$ to ${i-1}$, respectively.
This means that $t_{\rm rec}$ is inversely proportional on the electron density $n_{\rm e}$.
Photoionization plasma codes, such as {\tt SPEX}, calculate $t_{\rm rec}$ of any ion for a given $\xi$ and $n_{\rm e}$.
The relation between $n_{\rm H}$ and $n_{\rm e}$ is also computed by the photoionization code.
Therefore, if $t_{\rm rec}$ is measured from observations, the density $n_{\rm H}$ will be obtained, hence allowing the calculation of $r$ and $P_{\rm kin}$ of the wind.

%
\begin{figure*}
\centering
\resizebox{0.7\hsize}{!}{\includegraphics[angle=0]{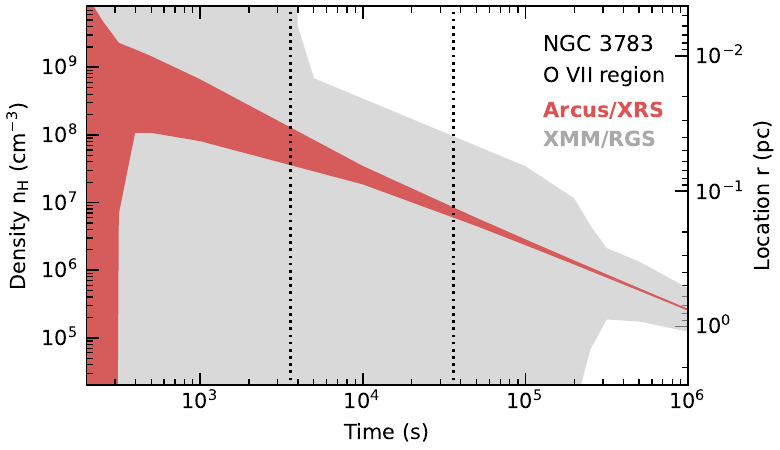}}
\caption{\small Constraints on the sought-after density and location of AGN outflows using the recombination timing of the O VII absorber in NGC 3783. The simulations with Arcus/XRS and XMM/RGS are compared at exposure times corresponding to the recombination time displayed on the x-axis. Shorter recombination and exposure times probe higher density gas closer to the black hole. The simulations show that with XMM/RGS we can only probe lower-density gas by resorting to extreme exposure times. However, Arcus enables us to probe a wide range of densities, including higher density gas at timescales of hours, thus allowing us to map different regions and components of the warm-absorber outflows. The vertical dotted lines indicate the region where most of the ionized gas in NGC 3783 is expected to reside, which can only be investigated with Arcus.
\label{fig_density}}
\vspace{-0.0cm}
\end{figure*}

The variability of the ionizing SED causes changes in the ionization/recombination in the gas, and hence in the ionic column densities, which translates into changes in the equivalent width of the observed absorption lines. 
By detecting such changes in the absorption lines in response to the SED, $t_{\rm rec}$ and thus $n_{H}$ are derived.
This means exposure time of the observation needs to be ${\ge t_{\rm rec}}$ in order to detect such line changes.
Because ${t_{\rm rec} \propto 1 / n_{\rm H}}$, as density becomes larger (i.e. $r$ becomes smaller) $t_{\rm rec}$ becomes shorter.
Therefore, with current instruments probing higher density outflows, closer to the black hole, becomes more challenging as the S/N per time bin becomes too low to apply this technique.
We demonstrate this point via simulations that we have carried out for the case of NGC 3783, shown in Fig. \ref{fig_density}.
Time-resolved spectroscopy of the X-ray absorption lines is currently not feasible with XMM/RGS, and we can only probe lower-density gas by resorting to extremely large exposure times.
However, the proposed Arcus mission overcomes this difficulty, as seen in Fig. \ref{fig_density}.
It allows us to trace changes in the absorption lines that occur on hours timescales, and hence probe outflows that are closer to the black hole.
%

%
\begin{figure*}
\centering
\resizebox{0.65\hsize}{!}{\includegraphics[angle=0]{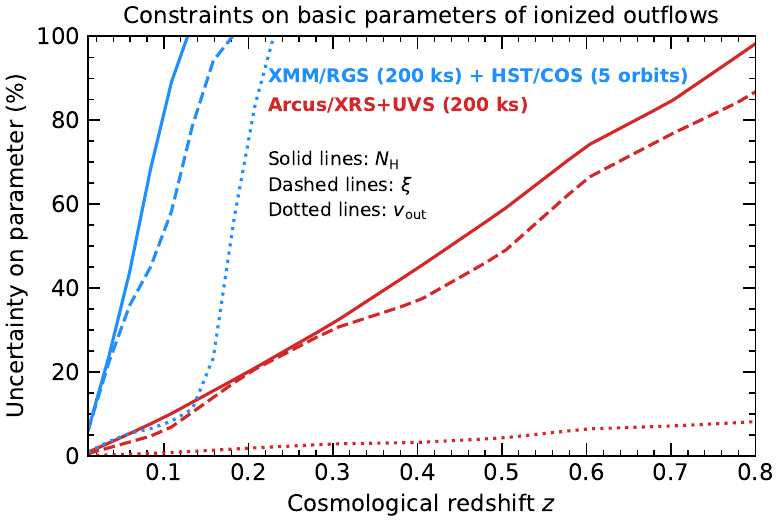}}
\caption{\small Comparison of constraints on the basic parameters of the warm-absorber outflows using simultaneous X-ray and UV spectroscopy with Arcus versus joint XMM/RGS and HST/COS spectroscopy. The simulations are carried out for an AGN with a moderate luminosity and a typical warm absorber outflow (described in Sect. \ref{sect_redshift}) observed at different cosmological redshifts $z$. The figure shows that with current missions, only an extremely limited redshift range in the local universe can be probed. However, Arcus not only measures parameters in this range very well (few \% uncertainty), it also pushes the limit for constraining the parameters to redshifts of about 0.8. This allows us to investigate how AGN wind properties scale from the local universe to such intermediate redshifts.
\label{fig_redshift}}
\vspace{-0.0cm}
\end{figure*}

\section{Pushing the frontiers of high-resolution spectroscopy beyond the local universe}
\label{sect_redshift}

The X-ray studies of AGN outflows are currently limited to the brightest objects in the local universe. 
This is because AGN are often too faint in X-rays for high-resolution spectroscopy with the existing missions.
In addition, with UV alone we cannot measure all components of the outflows, since the majority of the ionized gas appears in the X-ray band.
Therefore, we are currently restricted in our sample studies of outflows.
Importantly, Arcus enables us to probe AGN at higher redshifts to determine how winds operate beyond the nearest local universe.
Figure \ref{fig_redshift} demonstrates this point. 
The simulations are carried out for an AGN with a moderate bolometric luminosity (${3 \times 10^{44}}$ erg~s$^{-1}$, based on the SED of NGC 5548 \cite{Meh15a}), and a warm-absorber outflow with typical parameters: $\NH = 1 \times 10^{21}$~cm$^{-2}$, ${\log \xi = 2}$, and an outflow velocity of $500$~km~s$^{-1}$.

Using Arcus spectroscopy we can probe redshifts ${z \lesssim 0.8}$ (Fig. \ref{fig_redshift}). This opens up new AGN targets (and parameter space) that are currently not accessible with high-resolution spectroscopy with XMM/RGS: e.g. the Supermassive Black Hole Winds in X-rays (SUBWAYS) sample\cite{Matz23,Mehd23} at redshifts ${0.1 \lesssim z \lesssim 0.4}$.
To this end, Arcus plans to observe two samples of AGN: a Broad Sample for which the column density, ionization parameter, and velocity of the outflows are measured and their relation to the accretion rate is investigated; and a Deep Sample for which the time-dependent properties of the outflows will be measured with longer exposures, providing us with the sought-after density, location, and kinetic power of the outflows.
This expansion in wind studies by Arcus, probing a larger population of AGN, helps us in better understanding what physical factors govern the launch and duty cycle of AGN winds at different redshifts, and how the wind parameters scale with the accretion properties of the AGN.

\section{Conclusions}
\label{sect_concl}

Outflows in AGN consist of multiple components, with complex ionization and velocity structure.
Density is a particularly uncertain and important parameter of the ionized outflows, and holds clues to other other key questions, such as the origin and kinetic power of the outflows.
Simultaneous X-ray and UV spectroscopy and variability analysis are key for probing these poorly understood properties of AGN outflows.
This would become possible with the proposed Arcus mission, overcoming current limitations with existing missions.
Arcus would enable us to establish the ionization structure, kinematics, and energetics of various types of outflows, from transient obscuring disk winds in the BLR, to the warm-absorber outflows in the NLR.
Probing AGN outflows with Arcus in a much larger population than currently feasible, will help us in testing different theoretical models of outflows, and will broaden our understanding of how outflows in AGN are governed, and how they impact their host galaxies.

\subsection*{Disclosures}
The authors declare there are no financial interests, commercial affiliations, or other potential conflicts of interest that have influenced the objectivity of this research or the writing of this paper

\section* {Code and Data Availability}
The SPEX code v3.07.01 has been used for all our calculations in this paper. This code is publicly available at \url{https://doi.org/10.5281/zenodo.7037609}

\section* {Acknowledgments}
Support from NASA, STScI, SAO, MIT, and SRON are acknowledged.


\bibliography{references}   
\bibliographystyle{spiejour}   


\vspace{2ex}\noindent 
Dr. Missagh Mehdipour is currently a researcher at Space Telescope Science Institute. His research interests and experiences are on the astrophysics of AGN, namely the outflow and accretion/emission phenomena in these objects. He has expertise in X-ray and UV spectroscopic studies of the AGN outflows, broadband continuum modeling, and photoionization modeling. 
Dr. Laura Brenneman currently serves as Chair of the High-Energy Astrophysics Division at the CfA. Her research activities are: (1) characterizing outflowing winds in AGN and their influence on the structure and geometry of the galactic nucleus; (2) probing the inner accretion flows of supermassive black holes through X-ray spectroscopy.
Prof. Jon Miller is a professor of astronomy at the University of Michigan. He studies accretion onto black holes, and feedback between black holes and host galaxies primarily using X-ray, UV, and optical observations.
Dr. Elisa Costantini is a senior research scientist at SRON, Astro division. She is an expert in AGN winds physics, hot gas in the interstellar medium and interstellar dust. She is also an expert in laboratory astrophysics, especially related to interstellar dust.
Prof Ehud Behar is a high-energy astrophysicist, former dean of the Faculty of Physics, and the past director of the Norman and Helen Asher Space Research Institute (ASRI) at the Technion. His research expertise and interests are on X-ray spectroscopy of almost all types of astrophysical sources, as well as laboratory plasmas.
Dr. Luigi Gallo is a professor at Saint Mary’s University, Astronomy \& Physics Department. He is an expert on the topic of high-resolution spectroscopy of AGN. He has been involved in the development of several recent missions.
Prof. Jelle Kaastra is a senior scientist at SRON and a professor of high-energy Astrophysics at Leiden University. He is an expert on high-resolution X-ray spectroscopy using many missions. He is the main developer of the SPEX analysis package.
Dr. Sibasish Laha is a research scientist at NASA Goddard Space Flight Center. He is an expert in high-resolution X-ray spectroscopy (XMM-Newton and Chandra), studying outflows and their impact on the host galaxy. His research interests also include time-domain AGN astronomy (changing-look AGN and TDEs).
Dr. Michael Nowak is a research Professor at Washington University, Physics Department. His interests encompass observational and theoretical astrophysics of neutron stars, Galactic black holes, and AGN. Dr. Nowak is especially interested in astrophysical accretion processes.

\listoffigures
\listoftables

\end{spacing}
\end{document}